\documentstyle[11pt]{article}
\input amssymb.sty

\title{Interpreting the Modal Kochen-Specker Theorem:\\Possibility and Many Worlds\\in Quantum Mechanics}

\author{{\sc C. de Ronde}$^{*}$ $^{1,2}$, {\sc H. Freytes}\thanks{Fellow of the Consejo
Nacional de Investigaciones Cient\'{\i}ficas y T\'ecnicas
(CONICET). Tel. and email address of corresponding author: +54-11-4432-0606/ cderonde@vub.ac.be} \ 
$^{3,4}$ and  {\sc G. Domenech}  $^{2,5}$}
\date{}

\begin{document}

\bibliographystyle{plain}
\maketitle

\begin{center}
\begin{small}
1. Instituto de Filosof\'{\i}a  ``Dr. Alejandro Korn'' (UBA-CONICET)\\
2. Center Leo Apostel for Interdisciplinary Studies\\ Foundations of the Exact Sciences (Vrije Universiteit Brussel).\\
3. Dipartimento di Matematica e Informatica\\ ÒU. DiniÓ Viale Morgagni, 67/a - 50134 Firenze, Italia.\\
4. Departamento de matem‡tica (UNR-CONICET)\\Av Pellegrini 250 Rosario-Argentina.\\
5. Instituto de Astronom\'{\i}a y F\'{\i}sica del Espacio
(UBA-CONICET)

\end{small}
\end{center}

\begin{abstract}

\noindent In this paper we attempt to physically interpret the Modal
Kochen-Specker (MKS) theorem. In order to do so, we analyze the features
of the possible properties of quantum systems arising from the
elements in an orthomodular lattice and distinguish the use of
``possibility'' in the classical and quantum formalisms. Taking into account the 
 modal and many worlds non-collapse interpretation of the projection postulate, we discuss how the MKS theorem rules the constraints to actualization, and thus, the relation between actual and possible realms.
\end{abstract}

\begin{small}



\end{small}

\bibliography{pom}

\begin{thebibliography}{10}

\bibitem{Bub97} Bub, J., 1997, {\it Interpreting the
Quantum World}, Cambridge University Press, Cambridge.

\bibitem{Cohen}  Cohen-Tannoudji, C.,  Diu, B. and Lalo\"e, F., 1977, {\it Quantum Mechanics}, John Wiley and Sons, London.

\bibitem{DChGG} Dalla Chiara, M., Giuntini, R. and Greechie, R., 2004, {\it
Reasoning in Quantum Theory}, Kluwer Academic Publishers, Dordrecht.

\bibitem{DeWitt73} DeWitt, B., 1973, ``The Many-Universes
Interpretation of Quantum Mechanics'', In {\it Foundations of
Quantum Mechanics}, 167-218, Academic Press, New York.

\bibitem{DeWittGraham73} DeWitt, B. and Graham, N., 1973,
{\it The Many-Worlds Interpretation of Quantum Mechanics}, Princeton
University Press, Princeton.

\bibitem{Dickson} Dickson, W. M., 2001, ``Quantum logic is alive $\wedge$ (It is true $\vee$ It is false)'',
{\it Proceedings of the Philosophy of Science Association 2001},
{\bf 3}, S274-S287.

\bibitem{Dickson98} Dickson, W. M., 1998, {\it Quantum Chance
and Nonlocality: Probability and Nonlocality in the Interpretations
of Quantum Mechanics}, Cambridge University Press, Cambridge.

\bibitem{Dieks88a} Dieks, D., 1988, ``The Formalism of Quantum
Theory: An Objective Description of Reality'', {\it Annalen der
Physik}, {\bf 7}, 174-190.

\bibitem{Dieks89a} Dieks, D., 1989, ``Quantum Mechanics Without
the Projection Postulate and Its Realistic Interpretation'', {\it
Foundations of Physics}, {\bf 19}, 1397-1423.

\bibitem{Dieks07} Dieks, D., 2007, ``Probability in the modal Interpretation of quantum mechanics'', {\it Studies in
History and Philosophy of Modern Physics}, {\bf 38}, 292-310.

\bibitem{Dieks10} Dieks, D., 2010, ``Quantum Mechanics, Chance and
Modality", {\it Philosophica}, {\bf 83}, 117-137.

\bibitem{Dirac74} Dirac, P. A. M., 1974, {\it The Principles of
Quantum Mechanics}, 4th Edition, Oxford University Press, London.

\bibitem{DF} Domenech, G. and Freytes, H., 2005, ``Contextual
logic for quantum systems'', {\it Journal of Mathematical Physics},
{\bf 46}, 012102-1 - 012102-9.

\bibitem{DFR06} Domenech, G., Freytes, H. and de Ronde, C., 2006,
``Scopes and limits of modality in quantum mechanics",
\textit{Annalen der Physik}, {\bf 15}, 853-860.

\bibitem{DFR08} Domenech, G., Freytes, H. and de Ronde, C., 2008,
``A topological study of contextuality and modality in quantum
mechanics'', {\it International Journal of Theoretical Physics},
{\bf  47}, 168-174.

\bibitem{DFR09a} Domenech, G., Freytes, H. and de Ronde, C., 2009,
``Modal-type orthomodular logic'', {\it Mathematical Logic
Quarterly}, {\bf 3}, 307-319.

\bibitem{DFR09b} Domenech, G., Freytes, H. and de Ronde, C., 2009,
``Many worlds and modality in the interpretation of quantum
mechanics: an algebraic approach'', {\it Journal of Mathematical
Physics}, {\bf 50}, 072108.

\bibitem{Isham} D\"{o}ring, A. and Isham, C. J., 2008, ``A topos foundation
for theories of physics: I. Formal languages for physics'', {\it
Journal of Mathematical Physics}, {\bf 49}, 053515.

\bibitem{DI12} D\"{o}ring, A. and Isham, C. J., 2012, ``Classical and quantum probabilities as truth values'', {\it
Journal of Mathematical Physics}, {\bf 53}, 032101.

\bibitem{Everett57} Everett, H., 1957, ```Relative State'
Formulation of Quantum Mechanics'', {\it Reviews of Modern Physics},
{\bf 29}, 454-462.

\bibitem{Everett73} Everett, H., 1973, ``The Theory of the
Universal Wave Function'', In {\it The Many-Worlds Interpretation of
Quantum Mechanics}, DeWitt and Graham (Eds.), Princeton University
Press, Princeton.

\bibitem{GRW} Ghirardi, G. C. Rimini A. and Weber, T., 1986, ``Unified
Dynamics for Microscopic and Macroscopic Systems'', {\it Physical
Review D}, {\bf 34}, 470-491.

\bibitem{Heis73} Heisenberg, W., 1973, ``Development of Concepts in
the History of Quantum Theory", In {\it The Physicist's Conception
of Nature}, 264-275, J. Mehra (Ed.), Reidel, Dordrecht.

\bibitem{Ka}  Kalman, J. A., 1958, ``Lattices with involution'', {\it Transactions of the American Mathematical Society}  {\bf 87}, 485-491.

\bibitem{KAL}  Kalmbach, G., 1983,  {\it Ortomodular Lattices}, Academic Press, London.

\bibitem{KS} Kochen, S. and Specker, E., 1967, ``On the problem
of Hidden Variables in Quantum Mechanics", {\it Journal of
Mathematics and Mechanics}, {\bf 17}, 59-87. Reprinted in Hooker,
1975, 293-328.

\bibitem{Lewis86} Lewis, D., 1986, {\it On the Plurality of Worlds}, Blackwell Publishers, Harvard.

\bibitem{MM} Maeda, F. and  Maeda, S., 1970,  {\it Theory
of Symmetric Lattices}, Springer-Verlag, Berlin.

\bibitem{Redei01} R\'{e}dei, M., 2001, ``Von Neumann's concept of quantum logic and quantum
probability'', in {\it John von Neumann and the Foundations of Quantum Physics}, M. R\'{e}dei and M. St\"{o}tzner (Eds.), 153-172, Kluwer Academic Publishers,
Dordrecht.

\bibitem{deRonde11} de Ronde, C., 2011, {\it The Contextual and Modal Character of Quantum Mechanics: A Formal and Philosophical Analysis in the Foundations of Physics}, PhD dissertation, Utrecht University.

\bibitem{Sakurai} Sakurai, J. J. and Napolitano, J.,  2010, {\it Modern Quantum Mechanics}, Addison-Wesley, London.

\bibitem{VF81} Van Fraassen, B. C., 1981, ``A modal Interpretation of
Quantum Mechanics'', in {\it Current Issues in Quantum Logic},
229-258, E. G. Beltrametti and B. C. van Fraassen (Eds.),  Plenum,
New York.

\bibitem{VF91} Van Fraassen, B. C., 1991, {\it Quantum Mechanics: An
Empiricist View}, Clarendon, Oxford.

\bibitem{Vermaas99a} Vermaas, P. E., 1999, {\it A Philosophers
Understanding of Quantum Mechanics}, Cambridge University Press,
Cambridge.

\bibitem{VermaasDieks95} Vermaas, P. E. and Dieks, D., 1995, ``The
Modal Interpretation of Quantum Mechanics and Its Generalization to
Density Operators'', {\it Foundations of Physics}, {\bf 25},
145-158.

\bibitem{VN} Von Neumann, J., 1996, {\it Mathematical Foundations
of Quantum Mechanics}, Princeton University Press (12th. edition),
Princeton.

\end{thebibliography}

\newtheorem{theo}{Theorem}[section]

\newtheorem{definition}[theo]{Definition}

\newtheorem{lem}[theo]{Lemma}

\newtheorem{met}[theo]{Method}

\newtheorem{prop}[theo]{Proposition}

\newtheorem{coro}[theo]{Corollary}

\newtheorem{exam}[theo]{Example}

\newtheorem{rema}[theo]{Remark}{\hspace*{4mm}}

\newtheorem{example}[theo]{Example}

\newcommand{\proof}{\noindent {\em Proof:\/}{\hspace*{4mm}}}

\newcommand{\qed}{\hfill$\Box$}

\newcommand{\ninv}{\mathord{\sim}} 

\section*{Introduction}

In classical physics, every physical system may be described
exclusively by means of its \emph{actual properties}, taking
`actuality' as expressing the \emph{preexistent} mode of being of
the properties themselves, independently of observation ---the `pre'
referring to its existence previous to measurement. The evolution of
the system may be described by the change of its actual properties.
Mathematically, the state is represented by a point $(p,\ q)$ in the
corresponding phase space $\Gamma$ and, given the initial conditions, the equation of motion tells
us how this point moves in $\Gamma$.\footnote{For simplicity, we have in mind a system that is
only a material point.} Physical magnitudes are represented by real
functions over $\Gamma$. These functions commute with each other
and can be all interpreted as possessing definite values at any
time, independently of physical observation. In this scheme, speaking about potential or possible
properties usually refers to functions defined on points in $\Gamma$ to
which the state of the system will arrive at a future instant of
time; these points, in turn are completely determined by the equations of
motion and the initial conditions. 

In the orthodox formulation of quantum mechanics (QM), the
representation of the state of a system is given by a ray in Hilbert
space ${\cal H}$. Contrary to the classical scheme, physical
magnitudes are represented by operators on ${\cal H}$ that, in
general, do not commute. This mathematical fact has extremely
problematic interpretational consequences for it is then difficult
to affirm that these quantum magnitudes are \emph{simultaneously
preexistent}. In order to restrict the discourse to  sets of
commuting magnitudes, different Complete Sets of Commuting Operators
(CSCO) have to be chosen. This choice has not found until today a
clear justification and remains problematic. In the literature this feature is called
 {\it quantum contextuality}  ---it will be discussed
in section 1. Another fundamental feature of QM is
due to the linearity of the Schr\"{o}dinger equation which implies
the existence of entangled states involving the measuring device. The path from such an entangled state, i.e. a superposition of eigenstates of the measured observable to the eigenstate
corresponding to the measured eigenvalue is given, formally, by an
axiom added to the formalism: \emph{the projection postulate}. In
section 2 we will discuss the different physical interpretations of
this postulate which is, either thought in terms of a ``collapse'' of the wave
function (i.e., as a real physical interaction) or in terms of
non-collapse proposals, such as the modal and many worlds
interpretations. After having introduced and discussed these two
main features of QM we will present, in section 3,  our formal
analysis regarding possibility in orthomodular structures. In
section 4, we shall discuss and analyze the distinction between
mathematical formalism and physical interpretation, a distinction
which can raise many pseudo-problems if not carefully taken into
account. As a consequence of this distinction we will also put forward
the difference between `classical possibility' and `quantum
possibility'. In section 5, we are ready to advance towards a
physical interpretation of both quantum possibility and the MKS
theorem ---taking into account the specific formal constraints to modality
implied by it. In section 6 we will discuss the consequences of the
MKS theorem regarding the many worlds interpretation. Finally, in
section 7, we provide the conclusions of our work.

\section{Quantum Contextuality and Modality}

The idea that a {\it preexistent}  set of definite properties constitutes or describes
reality is one of the basic ideas which remains the fundament of all
classical physical theories and determines the possibility to
speak about an independent objective world, a world which does not
depend on our choices or consciousness. Physical reality can be then
conceived and analyzed in terms of a theory ---which describes a
preexistent world--- independently of actual observation. But, as it
is well known, this description of physical reality faces several
difficulties when presupposed  in the interpretation of the
quantum formalism.  In formal terms, this is demonstrated by the
Kochen-Specker (KS) theorem, which states that if we consider three
physical magnitudes represented by operators {\bf A}, {\bf B} and
{\bf C}, with {\bf A} commuting with {\bf B} and {\bf C} but {\bf B}
non-commuting with {\bf C}, the value of {\bf A} depends on the
choice of the context of inquiry; i.e. whether {\bf A} is considered
together with {\bf B} or together with {\bf C} \cite{KS}. From an
operational point of view, this is bypassed by considering the
context (in KS sense) as the experimental arrangement ---in line with the original
idea of N. Bohr. However, if we attempt to go beyond the discourse regarding
measurement results and provide some kind of realist representation
of what is going on according to QM, we need to
make sense of the indeterminateness of properties.
As Chris Isham and Andreas D\"{o}ring clearly point out:

\begin{quotation}
{\small ``When dealing with a closed system, what is needed is a
realist interpretation of the theory, not one that is
instrumentalist. The exact meaning of `realist' is infinitely
debatable but, when used by physicists, it typically means the
following:

\begin{enumerate}
\item The idea of `a property of the system' (i.e., `the value of a
physical quantity') is meaningful, and representable in the theory.

\item Propositions about the system are handled using Boolean
logic. This requirement is compelling in so far as we humans think
in a Boolean way.

\item There is a space of `microstates' such that specifying a
microstate leads to unequivocal truth values for all propositions
about the system. The existence of such a state space is a natural
way of ensuring that the first two requirements are satisfied.
\end{enumerate}

The standard interpretation of classical physics satisfies these
requirements, and provides the paradigmatic example of a realist
philosophy in science. On the other hand, the existence of such an
interpretation in quantum theory is foiled by the famous
Kochen-Specker theorem.'' \cite[p. 2]{Isham}}
\end{quotation}

Contextuality can be directly related to the impossibility to
represent a piece of the world as constituted by a set of definite
valued properties independently of the choice of the context. This
definition makes reference only to the actual realm. But as we know,
QM makes probabilistic assertions about measurement results.
Therefore, it seems natural to assume that QM does not only deal
with actualities but also with possibilities. Then the question
arises whether the space of possibilities is subject to the same
restrictions as the space of actualities. Formally, on the one hand,
the set of actualities is structured as the orthomodular lattice of
subspaces of the Hilbert space of the states of the system and, as
Michael Dickson remarks in \cite{Dickson}, the KS theorem (i.e., the
absence of a family of compatible valuations from subalgebras of the
orthomodular lattice to the Boolean algebra of two elements {\bf 2})
can be understood as a consequence of the failure of the distributive law in the lattice. 
On the other hand, given an adequate definition of the possibility operator
$\Diamond$ ---as the one developed in \cite{DFR06, DFR09a}---
the set of possibilities is the center of an enlarged structure. Since the elements of
the center of a structure are those which commute with all other
elements, one might think that the possible propositions defined in
this way escape from the constraints arising from the non-commutative
character of the algebra of operators. Thus, at first sight one
might assume that possibilities behave in a classical manner.

When predicting measuring results the context has been already
fixed. However, probability is a measure over the whole lattice and,
consequently,  the set of events over which the measure is defined
is non-distributive, calling attention to the interpretation of
possibility and probability. As noticed by Schr\"odinger in a letter
to Einstein \cite[p. 115]{Bub97}: ``It seems to me that the concept
of probability is terribly mishandled these days. Probability surely
has as its substance a statement as to whether something {\it is} or
{\it is not} the case --of an uncertain statement, to be sure. But
nevertheless it has meaning only if one is indeed convinced that the
something in question quite definitely {\it is} or {\it is not} the
case. A probabilistic assertion presupposes the full reality of its
subject.'' Also von Neumann was worried about a sound definition of
probability, as mentioned in \cite{Redei01}.\footnote{As
R\'{e}dei \cite[p. 157]{Redei01} states: ``To see why von Neumann insisted on the
modularity of quantum logic, one has to understand that he wanted
quantum logic to be not only the propositional calculus of a quantum
mechanical system but also wanted it to serve as the event structure
in the sense of probability theory. In other words, what von Neumann
aimed at was establishing the quantum analogue of the classical
situation, where a Boolean algebra can be interpreted both as the
Tarski-Lindenbaum algebra of a classical propositional logic and as
the algebraic structure representing the random events of a
classical probability theory, with probability being an additive
normalized measure on the Boolean algebra.''}
The difficulties with a rigorous definition of probability made von
Neumann abandon the orthodox formalism of QM in Hilbert space to which
he himself had so much contributed and face the
classification of the factors and their dimension functions which
led to the subject of von Neumann's algebras.\footnote{It might be
argued that a complete theory of quantum probability is still
lacking. On the one hand, type II$_1$ factor (the one whose
projection lattice is a continuous geometry, and thus an othomodular
modular lattice as required by a definition of probability) is not
an adequate structure to represent quantum events. On the other
hand, there exist different candidates for defining conditional
probability and there is not a unique criterium for choosing among
them \cite{DChGG}. Moreover, there are situations in which the
frequentist interpretation does not apply and consequently it is
required to develop new probability structures to account for
quantum phenomena \cite{DI12}.}

In order to explicitly verify whether modal propositions escape from
KS-type contradictions, in previous works we have developed a
mathematical scheme which allowed us to deal with both actual and
possible propositions in the same structure.\footnote{Van Fraassen
distinguishes two different isomorphic structures for dealing with
possible and actual properties (\cite{VF91}, chapter 9). The main
aspects of van Fraassen's modal interpretation in terms of quantum
logic  are as follows. The probabilities are of events, each
describable as `an observable having a certain value', corresponding
to value states. If $w$ is a physical situation in which system $X$
exists, then $X$ has both a {\it dynamic state} $\varphi$ and a {\it
value state} $\lambda$, i.e. $w=<\varphi, \lambda >$. A {\it value
state} $\lambda$ is a map of observable ${\bf A}$ into non-empty
Borel sets $\sigma$ such that it assigns \{1\} to $1_{\sigma}{\bf
A}$. $1_{\sigma}$ is the characteristic function of the set $\sigma$
of values. So, if the observable $1_{\sigma}{\bf A}$ has value 1,
then it is impossible that ${\bf A}$ has a value outside $\sigma$.
The proposition $<{\bf A}, \sigma> = \{ w:\lambda (w)({\bf
A})\subseteq \sigma \}$ assigns values to physical magnitudes, it is
a {\it value-attribution proposition} and  is read as `${\bf A}$
(actually) has value in $\sigma$'. $ {\cal V}$ is called the set of
value attributions ${\cal V}=\{ <{\bf A}, \sigma>:{\bf A}\ {\rm an \
observable \ and}\ \sigma \ {\rm a\ Borel\ set}\}$. The logic
operations among value-attribution propositions are defined as:
$<{\bf A}, \sigma>^{\bot} = <{\bf A}, \Re - \sigma>$, $<{\bf A},
\sigma>\wedge <{\bf A}, \theta>= <{\bf A}, \sigma\cap \theta>$,
$<{\bf A}, \sigma>\vee <{\bf A}, \theta>= <{\bf A}, \sigma\cup
\theta>$ and $\wedge\{ <{\bf A}, \sigma_{i}
>: i \in {\cal N}\} = <{\bf A}, \cap\{ \sigma_{i}: i \in {\cal
N}\}>$. With all this, ${\cal V}$ is the union of a family of
Boolean $\sigma$-algebras $<{\bf A}>$ with common unit and zero
equal to $<{\bf A}, S({\bf A})>$ and $<{\bf A}, \wedge >$
respectively. The Law of Excluded Middle is satisfied: every
situation $w$ belongs to $q\vee q^{\bot}$, but not the Law of
Bivalence: situation $w$ may belong neither to $q$ nor to
$q^{\bot}$. A {\it dynamic state} $\varphi$ is a function from
${\cal V}$ into [0, 1], whose restriction to each Boolean
$\sigma$-algebra $<{\bf A}>$ is a probability measure. The relation
between dynamic and value states is the following: $\varphi$ and
$\lambda$ are a dynamic state and a value state respectively, only
if there exist possible situations $w$ and $w'$ such that $\varphi
=\varphi (w),\ \lambda=\lambda (w')$. Here, $\varphi$ is an
eigenstate of ${\bf A}$, with corresponding eigenvalue $a$, exactly
if $\varphi(<{\bf A}, \{ a\}>)=1$. The {\it state-attribution
proposition} [${\bf A},\sigma$] is defined as: [${\bf A}, \sigma$] =
$\{ w: \varphi(w)(<{\bf A}, \sigma>) =1\}$ and means `${\bf A}$ must
have value in $\sigma$'. ${\cal P}$ denotes the set of
state-attribution propositions: ${\cal P}=\{ [{\bf A}, \sigma]: {\bf
A} \ {\rm an \ observable, }\ \sigma \ {\rm a \ Borel \ set}\}$.
Partial order between them is given by $[{\bf A}, \sigma ]\subseteq
[{\bf A}^{'} , \sigma^{'}]\ {\rm only\ if,\ for\ all\ dynamical\
states} \ \varphi, \ \varphi(<{\bf A}, \sigma>) \leq \varphi(<{\bf
A}^{'} , \sigma^{'}
>)$ and the logic operations are (well) defined as: $[{\bf A},
\sigma]^{\bot} = [{\bf A}, \Re -\sigma ]$, $[{\bf A}, \sigma]\uplus
[{\bf A}, \theta]= [{\bf A}, \sigma\cup \theta ]$ and $[{\bf A},
\sigma] \cap [{\bf A}, \theta] = [{\bf A}, \sigma\cap \theta]$. With
all this, $<{\cal P}, \subseteq , ^{\bot}>$ is an orthoposet, the
orthoposet formed by `pasting together' a family of Boolean algebras
in which whole operations coincide in areas of overlap. It may be
enriched to approach the lattice of subspaces of  Hilbert space.}
Within this frame we were able to prove a theorem which describes
the algebraic relations between both kinds of propositions. The
theorem shows explicitly the formal limits of {\it possible
actualizations}, in short, that no enrichment of the orthomodular
lattice with modal propositions allows us to circumvent the contextual
character of the quantum language. For obvious reasons, we called it
the Modal Kochen-Specker (MKS) theorem \cite{DFR06}. As in the case
of actual propositions, the MKS theorem may be demonstrated
with topological tools \cite{DFR08}. It is important to remark that
our formalism also provides a formal meaning in an algebraic frame
to the Born rule, something that has been discussed by Dieks in
relation to the possible derivation of a preferred probability
measure \cite{Dieks07}. Possible actualizations relate to the path
between the possible and the actual realms. As said before, within
QM this path remains extremely problematic in itself and needs to be
taken into account, contrary to classical physics, through the
introduction of an axiom, the projection postulate. In the following
section we will analyze its meaning and possible interpretation.

\section{Projection Postulate and Quantum Collapse}

Classical texts that describe QM axiomatically begin stating that
the mathematical interpretation of a quantum system is a Hilbert
space, that pure states are represented by rays in this space,
physical magnitudes by selfadjoint operators on the state space and
that the evolution of the system is ruled by the Schr\"{o}dinger
equation. Possible results of a given magnitude are the eigenvalues
of the corresponding operator obtained with probabilities given 
by the Born rule. In general the state previous to the
measurement is a linear superposition of eigenstates corresponding
to different eigenvalues of the measured observable. 
In order to give an account of the state of the system after the
appearance of a particular result a new axiom needs to be
added: the {\it projection postulate}.  In von Neunmann's \cite[p. 214]{VN} words:
``Therefore, if the system is initially found in a state in which
the values of $\mathcal{R}$ cannot be predicted with certainty, then
this state is transformed by a measurement $M$ of $\mathcal{R}$ into
another state: namely, into one in which the value of $\mathcal{R}$
is uniquely determined. Moreover, the new state, in which $M$ places
the system, depends not only on the arrangement of $M$, but also on
the result of $M$ (which could not be predicted causally in the
original state) ---because the value of $\mathcal{R}$ in the new
state must actually be equal to this $M$-result''.\footnote{Or in
Dirac's words: ``When we measure a real dynamical variable $\xi$,
the disturbance involved in the act of measurement causes a jump in
the state of the dynamical system. From physical continuity, if we
make a second measurement of the same dynamical variable $\xi$
immediately after the first, the result of the second measurement
must be the same as that of the first'' \cite[p. 36]{Dirac74}.} At
this point one needs to introduce the so called {\it
eigenstate-eigenvalue link: after the measurement, the state
of the system is that (i.e., the eigenstate) which corresponds to
the measured eigenvalue.}

There are different ways to account for the projection
postulate. One of them is to consider it as a ``collapse'' which
takes place during measurement, i.e. as a real physical stochastic
``jump'' from the state previous to the measurement to the
eigenstate corresponding to the measured eigenvalue \cite{Cohen,
Sakurai}. This interpretation was strongly debated by the founding
fathers, Schr\"{o}dinger himself is quoted to have said:  ``Had I
known that we were not going to get rid of this damned quantum
jumping,  I never would have involved myself in this
business!''.\footnote{As noticed by Dieks \cite[p. 120]{Dieks10}: ``Collapses constitute a
process of evolution that conflicts with the evolution governed by
the Schr\"{o}dinger equation. And this raises the question of
exactly when during the measurement process such a collapse could
take place or, in other words, of when the Schr\"{o}dinger equation
is suspended. This question has become very urgent in the last
couple of decades, during which sophisticated experiments have
clearly demonstrated that in interaction processes on the
sub-microscopic, microscopic and mesoscopic scales collapses are
never encountered.''}
Another way of dealing with the postulate is by adding non-linear terms to the equation of evolution that may conduce to the eigenstate in a stochastic manner, as in the case of GRW
theory \cite{GRW}.

A different way to approach the problem is to deny the existence of
a collapse during measurement but still use something like the projection postulate
as an interpretational rule. A well known non-collapse
interpretation is the so called modal interpretation (MI) of QM.
This approach states that superpositions remain always intact, independent of the result of the actual
observation.\footnote{Van Fraassen discusses the problems of the
collapse of the quantum wave function in \cite{VF91}, section 7.3.
See also \cite{Dickson98}. Dieks \cite[p. 182]{Dieks88a} argues
that: ``[...] there is no need for the projection postulate. On the
theoretical level the full superposition of states is always
maintained, and the time evolution is unitary. One could say that
the `projection' has been shifted from the level of the theoretical
formalism to the semantics: it is only the empirical interpretation
of the superposition that the component terms sometimes, and to some
extent, receive an independent status.''} One might
say that the eigenstate-eigenvalue link is here accepted only in one
direction, implying that given a state that is an eigenstate there
is a definite value of the corresponding magnitude, i.e. its
eigenvalue, but not the other way around. ``In modal interpretations the state
is not updated if a certain state of affairs becomes actual. The
non-actualized possibilities are not removed from the description of
a system and this state therefore codifies not only what is presently
actual but also what was presently possible. These non-actualized
possibilities can, as a consequence, in principle still affect the
course of later events." \cite[p. 295]{Vermaas99a} There are thus,
within MI, two independent levels given by the possible and the
actual.\footnote{These levels are explicitly formally accounted for
in both van Fraasen and Dieks MI. While van
Fraassen distinguishes between the `dynamical states' and the `value
states', Dieks and Vermaas consider a distinction between `physical
states' and `mathematical states' \cite{VermaasDieks95}.} The passage from the possible
realm to the actual realm is given through different
interpretational rules, depending on the version of the MI
\cite{Vermaas99a}. Anyway, since all possibilities in the superposition are retained, one could argue that within this scheme ---even though van Fraassen\footnote{According to van
Fraassen: modalities are in our theories, not in the world.} and Dieks have taken a stance in favor of an empiricist position
regarding modality \cite{VF81, Dieks10}--- there
is a place to interpret possibility in an ontological fashion
\cite{deRonde11}.

Many Worlds Interpretation (MWI) of QM is another well
known non-collapse interpretation which has become an
important line of investigation within the foundations of quantum
theory domain. It is considered to be a direct  conclusion from
Everett's first proposal in terms of `relative states'
\cite{Everett57}. Everett's idea was to let QM find
its own interpretation, doing justice to the symmetries inherent in
the Hilbert space formalism in a simple and convincing way
\cite{DeWittGraham73}. The solution proposed to the measurement
problem is that each one of the terms in the
superposition is {\it actual} in its own corresponding world. Thus,
it is not only the single value which we see in `our world' which
gets actualized but rather, that a branching of worlds takes place
in every measurement, giving rise to a multiplicity of worlds with
their corresponding actual values. The possible splits of the worlds
are determined by the laws of QM.

\begin{quotation}
{\small ``The whole issue of the transition from `possible' to
`actual' is taken care of in the theory in a very simple way
---there is no such transition, nor is such a transition necessary
for the theory to be in accord with our experience. {\small {\it
From the viewpoint of the theory all elements of a superposition
(all `branches') are `actual', none any more `real' than the rest.}}
It is unnecessary to suppose that all but one are somehow destroyed,
since all the separate elements of a superposition individually obey
the wave equation with complete indifference to the presence or
absence (`actuality' or not) of any other elements. This total lack
of effect of one branch on another also implies that no observer
will ever be aware of any `splitting' process." \cite[pp.
146-147]{Everett73}}\end{quotation}

\noindent In this case, there is no need to conceptually distinguish between
possible and actual because each state is actual within its own
branch and the eigenstate-eigenvalue link is maintained in each
world.

\section{On the Formal Limits of Possibility}

After having discussed some interpretational aspects of both modality and actualization we now shortly review our own development and analysis of the notion of possibility inside the formalism. First we recall from \cite{KAL, MM} some notions about orthomodular lattices.  A {\it lattice with involution} \cite{Ka} is an algebra
$\langle \mathcal{L}, \lor, \land, \neg \rangle$ such that $\langle
\mathcal{L}, \lor, \land \rangle$ is a  lattice and $\neg$ is a
unary operation on $\mathcal{L}$ that fulfills the following
conditions: $\neg \neg x = x$ and $\neg (x \lor y) = \neg x \land
\neg y$.  An {\it orthomodular lattice} is an algebra $\langle {\cal L},
\land, \lor, \neg, 0,1 \rangle$ $\leftarrow$  of type $\langle
2,2,1,0,0 \rangle$ that satisfies the following conditions:

\begin{enumerate}
\item
$\langle {\cal L}, \land, \lor, \neg, 0,1 \rangle$  is a bounded
lattice with involution,

\item
$x\land  \neg x = 0 $.

\item
$x\lor ( \neg x \land (x\lor y)) = x\lor y $

\end{enumerate}

We denote by ${\cal OML}$ the variety of orthomodular lattices. Let
$\mathcal{L}$ be an orthomodular lattice. {\it Boolean algebras} are
orthomodular lattices satisfying  the {\it distributive law} $x\land
(y \lor z) = (x \land y) \lor (x \land z)$. We denote by ${\bf 2}$
the Boolean algebra of two elements. Let $\mathcal{L}$ be an
orthomodular lattice. An element $c\in \mathcal{L}$ is said to be a
{\it complement} of $a$ iff $a\land c = 0$ and $a\lor c = 1$. Given
$a, b, c$ in $\mathcal{L}$, we write: $(a,b,c)D$\ \   iff $(a\lor
b)\land c = (a\land c)\lor (b\land c)$; $(a,b,c)D^{*}$ iff $(a\land
b)\lor c = (a\lor c)\land (b\lor c)$ and $(a,b,c)T$\ \ iff
$(a,b,c)D$, (a,b,c)$D^{*}$ hold for all permutations of $a, b, c$.
An element $z$ of $\mathcal{L}$ is called {\it central} iff for all
elements $a,b\in \mathcal{L}$ we have\ $(a,b,z)T$. We denote by
$Z(\mathcal{L})$ the set of all central elements of $\mathcal{L}$
and it is called the {\it center} of $\mathcal{L}$.

\begin{prop}\label{eqcentro} Let $\mathcal{L}$ be an orthomodular lattice. Then we have:

\begin{enumerate}

\item
$Z(\mathcal{L})$ is a Boolean sublattice of $\mathcal{L}$ {\rm
\cite[Theorem 4.15]{MM}}.

\item
$z \in Z(\mathcal{L})$ iff for each $a\in \mathcal{L}$, $a = (a\land
z) \lor (a \land \neg z)$  {\rm \cite[Lemma 29.9]{MM}}.

\end{enumerate}
\qed
\end{prop}

In the tradition of quantum logical research, a property of (or
a proposition about) a quantum system is related to a closed
subspace of the Hilbert space ${\mathcal H}$ of its (pure) states
or, analogously, to the projector operator onto that subspace.
Moreover, each projector is associated to a dichotomic question
about the actuality of the property \cite[p. 247]{VN}. A physical
magnitude ${\mathcal M}$ is represented by an operator ${\bf M}$
acting over the state space. For bounded self-adjoint operators,
conditions for the existence of the spectral decomposition ${\bf
M}=\sum_{i} a_i {\bf P}_i=\sum_{i} a_i |a_i\rangle\langle a_i|$ are
satisfied. The real numbers $a_i$ are related to the outcomes of
measurements of the magnitude ${\mathcal M}$ and projectors
$|a_i\rangle\langle a_i|$ to the mentioned properties. The physical
properties of the system are organized in the lattice of closed
subspaces ${\mathcal L}({\mathcal H})$ that, for the finite
dimensional case, is a modular lattice, and an orthomodular one in
the infinite case \cite{MM}. Moreover, each self-adjoint operator
$\bf M$ has an associated Boolean sublattice $W_{\bf{M}}$ of
${\mathcal L}({\mathcal H})$ which we will refer to as the spectral
algebra of the operator $\bf M$. Assigning values to a physical
quantity ${\cal M}$ is equivalent to establishing a Boolean
homomorphism $v: W_{\bf{M}} \rightarrow {\bf 2}$. As it is well
known, the KS theorem rules out the non-contextual assignment of
definite values to the physical properties of a quantum system. This
may be expressed in terms of valuations over  ${\mathcal
L}({\mathcal H})$ in the following manner. We first introduce the
concept of global valuation. Let  $(W_i)_{i\in I}$ be the family of
Boolean sublattices of ${\mathcal L}({\mathcal H})$. Then a {\it
global valuation} of the physical magnitudes over ${\mathcal
L}({\mathcal H})$ is a family of Boolean homomorphisms $(v_i: W_i
\rightarrow {\bf 2})_{i\in I}$ such that $v_i\mid W_i \cap W_j =
v_j\mid W_i \cap W_j$ for each $i,j \in I$. If this global valuation
existed, it would allow to give values to all magnitudes at the same
time maintaining a {\it compatibility condition} in the sense that
whenever two magnitudes shear one or more projectors, the values
assigned to those projectors are the same from every context. The KS
theorem, in the algebraic terms, rules out the existence of global
valuations when $dim({\mathcal H})>2$:

\begin{theo}\label{CS2} {\rm \cite[Theorem 3.2]{DF}}
If $\mathcal{H}$ is a Hilbert space such that $dim({\cal H}) > 2$,
then a global valuation, i.e. a family of Boolean homomorphisms over
the spectral algebras satisfying the compatibility condition, over
$L({\mathcal H})$ is not possible. \qed
\end{theo}

In what follows we delineate a modal extension for orthomodular
lattices that allows to formally represent, within the same
algebraic structure, actual and possible properties of the system.
This allows us to  discuss the restrictions posed by the theory
itself to the {\it actualization} of  possible properties. Given a
proposition about the system, it is possible to define a context
from which one can predicate with certainty about it together with a
set of propositions that are compatible with it and, at the same
time, predicate probabilities about the other ones (Born rule). In
other words, one may predicate truth or falsity of all possibilities
at the same time, i.e., possibilities allow an interpretation in a
Boolean algebra. In rigorous terms, let $P$ be a proposition about a
system and consider it as an element of an orthomodular lattice
${\cal L}$. If we refer with $\Diamond P$  to the possibility of $P$
then, by Proposition \ref{eqcentro}, we assume that $\Diamond P \in
Z({\cal L})$.

This interpretation of possibility in terms of the Boolean algebra
of central elements of ${\cal L}$ reflects the fact that one can
simultaneously predicate about all possibilities because Boolean
homomorphisms of the form $v:Z({\cal L}) \rightarrow {\bf 2}$ can be
always established. If $P$ is a proposition about the system and $P$
occurs, then it is trivially possible that $P$ occurs. This is
expressed as $P \leq \Diamond P$. Classical consequences that are
compatible with a given property, for example probability
assignments to the actuality of other propositions, share the
classical frame. These consequences are the same ones as those which
would be obtained by considering the original actual property as a
possible property. This is interpreted as, if $P$ is a property of
the system, $\Diamond P$ is the smallest central element greater
than $P$.

This enriched orthomodular structure can be axiomatized as a variety denoted by ${\cal OML}^\Diamond$ \cite[Theorem
4.5]{DFR06}. More precisely, each element of ${\cal OML}^\Diamond$
is an algebra $ \langle {\mathcal{L}}, \land, \lor, \neg, \Diamond,
0, 1 \rangle$ of type $ \langle 2, 2, 1, 1, 0, 0 \rangle$ such that
$ \langle {\mathcal{L}}, \land, \lor, \neg, 0, 1 \rangle$ is an
orthomodular lattice and $\Diamond$ satisfies the following
equations:

\begin{enumerate}

\item[S1]
$x \leq \Diamond x$ \hspace{3cm} S5 \hspace{0.2cm} $y = (y\land \Diamond x) \lor (y \land \neg \Diamond x)$

\item[S2]
$\Diamond 0 = 0$  \hspace{3.1 cm} S6 \hspace{0.2cm} $\Diamond (x \land \Diamond y ) = \Diamond x \land \Diamond y $

\item[S3]
$\Diamond \Diamond x = \Diamond x$  \hspace{2.5cm} S7 \hspace{0.2cm} $\neg \Diamond x \land \Diamond y   \leq \Diamond (\neg x \land (y \lor x)) $

\item[S4]
$\Diamond (x \lor y) = \Diamond x  \lor  \Diamond y$

\end{enumerate}

Each algebra of ${\cal OML}^\Diamond$ is called a {\it Boolean
saturated orthomodular lattice}. Orthomodular complete lattices are
examples of Boolean saturated orthomodular lattices. If ${\cal L}$
is a Boolean saturated orthomodular lattice, it is not very hard to
see that for each $x\in {\cal L}$, $$\Diamond x = Min \{z\in Z({\cal
L}): x\leq z  \}  $$ We can embed each orthomodular lattice
$\mathcal{L}$ in an element $\mathcal{L}^{\Diamond} \in  {\cal
OML}^\Diamond$ (see \rm \cite[Theorem 10]{DFR06}). In general,
$\mathcal{L}^{\Diamond}$ is referred to as a {\it modal extension of
$\mathcal{L}$}. This modal extension represents the fact that each
orthomodular system can be modally enriched in such a way as to
obtain a new propositional system that includes the original
propositions in addition to their possibilities. These possibilities
are formulated as classical propositions. Let $\mathcal{L}$ be an
orthomodular lattice and $\mathcal{L}^{\Diamond}$ a modal extension
of $\mathcal{L}$. We define the possibility space of $\mathcal{L}$
in $\mathcal{L}^{\Diamond}$ as as the subalgebra of
$\mathcal{L}^{\Diamond}$ generated by the set $\{\Diamond (P): P\in
{\cal L} \}$. This algebra is denoted by $\Diamond {\cal L}$ and we
can prove that  it is a Boolean subalgebra of the modal extension.

The possibility space represents the modal content added to the
discourse about properties of the system. Within this frame, the
actualization of a possible property acquires a rigorous meaning.
Let ${\cal L}$ be an orthomodular lattice, $(W_i)_{i \in I}$ the
family of Boolean sublattices of ${\cal L}$ and ${\cal L}^\Diamond$
a modal extension of $\cal L$. If $f: \Diamond {\cal L} \rightarrow
{\bf 2}$ is a Boolean homomorphism, an actualization compatible with
$f$ is a global valuation $(v_i: W_i \rightarrow {\bf 2})_{i\in I}$
such that $v_i\mid W_i \cap \Diamond {\cal L} = f\mid W_i \cap
\Diamond {\cal L} $ for each $i\in I$. A kind of converse of this
possibility of actualizing properties may be read as an algebraic
representation of the Born rule, something that has no place in the
orthomodular lattice alone. {\it Compatible actualizations}
represent the passage from possibility to actuality, they may be
regarded as formal constrains when applying the interpretational
rules proposed in the different modal versions. When taking into
account compatible actualizations from different contexts, an
analogon of the KS theorem holds for possible properties.

\begin{theo}\label{ksm}{\rm \cite[Theorem 6.2]{DFR06}}
Let $\cal L$ be an orthomodular lattice. Then $\cal L$ admits a
global valuation iff for each possibility space there exists a
Boolean homomorphism  $f: \Diamond {\cal L} \rightarrow {\bf 2}$
that admits  a compatible actualization.\qed
\end{theo}

\noindent The MKS theorem shows that no enrichment of the
orthomodular lattice with modal propositions allows us to circumvent
the contextual character of the quantum language. Thus, from a
formal perspective, one is forced to conclude that quantum
possibility is something different from classical possibility.

\section{Distinguishing the Mathematical Formalism from its Physical Interpretation}

The almost direct relation between classical logic and natural
language is not respected within QM. We argue that
this fact must be carefully taken into account and might be
responsible for pseudo problems when considering the question
``what is QM talking about?'' In the following
section, we attempt to provide a clear distinction between the
algebraic structure, its corresponding formal language and the
meta-language used in the theory. 

At this point, regarding the question of interpretation, we need to
be explicit about the stance we shall take regarding the possibility of going beyond the concepts of classical physics.
Following Dieks we argue that one should not demand that classical
physics should determine the conceptual tools of new theories, for
this ``would deny the possibility of really new fundamental
theories, conceptually independent of classical physics.''
\cite[p. 1417]{Dieks89a} Thus, we do not take for granted any self evident and univocal interpretation of a mathematical
formalism ---i.e., a pre-established set of concepts which have to
be necessarily applied to interpret mathematical structures. As
Heisenberg \cite[p. 264]{Heis73} remarked: ``The history of
physics is not only a sequence of experimental discoveries and
observations, followed by their mathematical description; it is also
a history of concepts. For an understanding of the phenomena the
first condition is the introduction of adequate concepts. Only with
the help of correct concepts can we really know what has been
observed.''

In mathematical terms, a context is a Boolean subalgebra of the
complete lattice. Thus, it may seem that the natural language we use
to refer to the compatible magnitudes represented by these commuting
operators poses no problem. However, this is not the case, due to the fact it is also necessary to take into account the state of the system (usually a superposition) when interpreting this algebra as the algebra of
a set of magnitudes of a physical system. Indeed, we have to
consider two very different cases: maybe the state is an eigenstate
of the CSCO ---a trivial case in which the values of all magnitudes
represented by the operators in the CSCO are determined, even when
not measured--- or it may be the case that the state of the system is not an
eigenstate ---the general case in which the CSCO
does not fix determinate values; in fact it only establishes which
magnitudes we are interested in.  In this case, when we refer to
possible properties we have to keep in mind that the meaning of
``posssible'' is not the same in the Boolean structure of classical
logic and in the Boolean subalgebra of the orthomodular structure of
QM. Let us make this point clear.  Note that if ${\cal L}$ is a
Boolean saturated orthomodular lattice in which ${\cal L}$ is a
Boolean algebra, $\Diamond$ is the identity operator. This can be
seen from the fact that,  if ${\cal L}$ is a Boolean structure,
${\cal L} = Z({\cal L})$  and then $\Diamond x = Min \{z\in Z({\cal
L}): x\leq z \} = x $ since $x\leq x$ and $x\in Z({\cal L})$. In
other words, the concept of possibility that corresponds to the
definition of the previous section becomes a ``trivial possibility''
in a classical structure. We have indicated with $\Diamond$ the
possibility operator related to the Boolean structure and we add a
subindex $Q$ for the quantum case. Thus, $\Diamond_Q$ is the
$\Diamond$ related to the orthomodular structure. Summing up,  \\

\begin{tiny}
\begin{center}
\begin{tabular}{|c|c|c|c|}

\hline
& \textbf{Algebraic Structure} & \textbf{Language} & \textbf{Meta-Language} \\
\hline
\textbf{Classical}  & \tiny{\textsl{Boolean lattice,}} & \tiny{{\it $\Diamond$ means possibility}} & \tiny{{\it (classical) possibility}}\\
\textbf{Mechanics}  & \tiny{\textsl{$\Diamond$}} & \tiny{{\it within a Boolean structure}} & \\
\hline
\textbf{Quantum}  & \tiny{\textsl{orthomodular lattice,}} & \tiny{{\it $\Diamond_{Q}$ means possibility}} & \tiny{{\it quantum possibility}}\\
\textbf{Mechanics}  & \tiny{\textsl{$\Diamond_{Q}$}} & \tiny{{\it within an orthomodular structure}} & \\
\hline
\end{tabular}
\end{center}\end{tiny}

\vspace{0.5cm}

In the classical case, the elements $A\in \wp (\Gamma )$ interpreted
as the properties of the system are part of a Boolean algebra (with
$\Gamma$ the classical phase space and $\wp (\Gamma)$ its power
set). The elements of the corresponding modal structure are
constructed by applying the possibility operator $\Diamond$ to the
elements $A$. These new elements $\Diamond A$, that belong to the
modal structure, correspond to possible properties as formulated in the
natural language. However, in this case, the seemingly larger
structure that includes both actual and modal propositions does not
enlarge the expressive power of the language. This is due to the
fact that there exists a trivial correspondence between any pair of
classical  valuations $v_{c}$ and $w_c$  of the actual and the
possible structures to truth-falsity. This relation can be written
as follows: let $A_k\in\wp(\Gamma)$, $k$ a fix index, then:
$$w_c(\Diamond A_k) = 1 \Leftrightarrow v_c(A_k) = 1$$
$$w_c(\Diamond A_k) = 0 \Leftrightarrow v_c(A_k) = 0$$

\noindent Thus, given the state of a classical system, possible properties at
a certain time coincide with (simultaneous) actual ones, they may be
identified. The distinction between the two sets of properties
need never be made. In fact, when referring to possible properties of a classical
system in a given state, one is always making reference to
\emph{future} possible values of the magnitudes, values that are determined because they are the evaluation of
functions at points $(p, \ q)$ in $\Gamma$ at  future times. These points
are determined in turn by the equation of motion. Thus, not even future
possibilities are classically indeterminate and they coincide with
\emph{future actual properties}.

On the contrary, in the quantum case, the projectors ${\bf
P}_a=|a\rangle\langle a|$ on $\mathcal{H} $, which are interpreted
as the properties of a system, belong to an orthomodular structure.
As we have mentioned above, the orthomodular lattice is enlarged with its modal content by
adding the elements $\Diamond_{Q} |a\rangle\langle a|$. Due to the
fact that there is no trivial relation between the valuations of
subsets of the possible and actual elements to truth-falsity, this
new structure  genuinely enlarges the expressive power of the
language. Formally, if $w_q(\Diamond_{Q} {\bf P}_k) = 1$, with ${\bf
P}_k\in W_i$, then there exists a valuation $v_q$ such that
$v_q({\bf P}_k) = 1$ and another $v'_q$ such that $v'_q({\bf P}_k) =
0$. Thus, contrary to the classical case, even at the same instant
of time, we may consider two different kind of properties, two
different realms, possible and actual, that do not coincide.
In order not to misinterpret the $\Diamond_{Q}$
operator, it is of great importance to clearly distinguish between
the formal language and the metalanguage. As a matter of fact, both
$\Diamond A$ and $\Diamond_{Q} |a\rangle\langle a|$ are called {\it
possible} within their own structures even though, at least
formally, the meaning of ``possible'' in each case is different.

\section{Quantum Possibility and the Physical Interpretation of the MKS Theorem}

In the literature regarding QM many times the
classical notion of possibility is silently assumed as a tool to interpret the formalism.
As we have argued above, there is however no reason why such
interpretation of the formalism should be necessarily applied, 
rather, this is part of an interpretational choice. In this
section, we are mainly interested in the physical interpretation of
the MKS theorem and the consequences and constraints it might
determine, within the formalism, for applying a coherent
interpretation to $\Diamond_{Q} |a\rangle\langle a|$. For this
purpose we shall explicitly distinguish between these two notions of
possibility, both formally and linguistically. To avoid any
misunderstanding we shall use ``possibility'' in relation to the
operator appearing in any Boolean structure ($\Diamond$) and ``quantum
possibility'' when the operator relates  to an orthomodular
structure ($\Diamond_{Q}$).

The distance between quantum and classical possibilities is also related to the formal difference between classical and quantum probabilities. The possibility of actualization of a physical property in classical statistical mechanics is given by the {\it probability weights}. In this case the probability is Kolmogorovian and can be interpreted as epistemic; i. e. as providing information of an unknown ---but existent--- state of affairs. In the quantum case, the wave function is a linear combination of vectors (each one in correspondence with a projector in the orthomodular lattice, associated to a physical property) with complex
coefficients interpreted as {\it probability amplitudes}. Contrary to the classical case, the probability implied by this structure is a non-Kolmogorovian one, and cannot be interpreted in terms of ignorance. As one symptom, when evaluating quantum probabilities there is an {\it interference term} which does not appear within the classical probability scheme. So, at least formally, probabilities in QM interfere.\footnote{As noticed by Dieks \cite[p. 124-125]{Dieks10}: ``In classical physics the most fundamental description of a physical system (a point in phase space) reflects only the actual,
and nothing that is merely possible. It is true that sometimes
states involving probabilities occur in classical physics: think of
the probability distributions $\rho$ in statistical mechanics. But
the occurrence of possibilities in such cases merely reflects our
ignorance about what is actual. The statistical states do not
correspond to features of the actual system (unlike the case of the
quantum mechanical superpositions), but quantify our lack of
knowledge of those actual features. This relates to the essential
point of difference between quantum mechanics and classical
mechanics [...]: in quantum mechanics the
possibilities contained in the superposition state may interfere
with each other. There is nothing comparable in classical physics.
In statistical mechanics the possibilities contained in $\rho$
evolve separately from each other and do not have any mutual
influence. Only one of these possibilities corresponds to the actual
situation."}

In order to physically interpret our MKS theorem what we need is an explicit map between the formal language and the meta-language. Instead of presupposing a set of metaphysical principles from which the formalism needs to be developed, our proposal attempts to provide a coherent interpretation starting from what we know about the formalism itself, and the structures it determines. In order to do so, we construct a dictionary relating names to the elements of the different structures:

\begin{enumerate}
\item $\Diamond A_i$ with $A_i$ in the Boolean lattice $\wp(\Gamma)$
is called ``possibility of $A_i$''.
\item $\Diamond_{Q} {\bf
P}_i$ with ${\bf P}_i$ in the orthomodular lattice $\mathcal{L} $ is
called ``quantum possibility of ${\bf P}_i$''.
\item The set of all the  $\Diamond_{Q} {\bf
P}_i$ with ${\bf P}_i$ in the orthomodular lattice $\mathcal{L} $ is
called the ``set of quantum possibilities''.
\item A Boolean sub-algebra of the orthomodular lattice $\mathcal{L} $ is called a ``context''.
\item The set of quantum possibilities valuated to $1\in {\bf 2}$ is called the ``set of existent quantum possibilities''.
\item The subset of quantum possibilities in direct relation to a context valuated to $1\in {\bf 2}$  is
called the ``set of existent quantum possibilities in a situation''.
\item The subset of projectors of the context  valuated to $1\in {\bf 2}$ is called the ``actual state of
affairs''.
\end{enumerate}

\noindent Physically, it follows from the given definitions that:

\begin{enumerate}
\item An ``actual state of affairs''
provides a physical description in terms of definite valued
properties.
\item A ``situation'' provides a physical description in terms
of the quantum possibilities that relate to an actual state of
affairs.
\item Formally, to go from the ``set of existent quantum possibilities''  to one of its
subsets (each of which relates to a ``context'') is to define an
application; physically, this path relates to the choice of a
particular measurement set up, restricting the expressiveness of the total set of existent possibilities to a specific subset.
\item Formally, to give values to the projectors ${\bf
P}_i$ in a context is to valuate; physically, the valuation
determines the set of properties (in correspondence with the
projectors ${\bf P}_i$ valuated to $1\in {\bf 2}$) which are
considered as preexistent.
\item The ``situation'' expresses an existent set of quantum possibilities
(which must not be considered in terms of actuality) while the
valuated context  expresses an actual state of affairs. This leaves
open the opportunity to consider quantum possibility as determining
a different mode of existence (independent of that of actuality).
\end{enumerate}

\noindent We have distinguished between a `situation' which makes reference to a definite set of existent possibilities and an `actual state of affairs' which can be interpreted as a specific measurement set up.
The `context' is the limit in between the actual and the possible and makes reference to the
non-commutative formal structure. As we have argued above it is important to notice this distinction is made relative to a single instant of time. It is the MKS theorem, understood within this specific scheme, which exposes the fact that the actualization of possible properties cannot be understood in terms of a classical path. Forcing the classical notion of possibility within the quantum structure is a move that contradicts the mathematical formalism, which contemplates the interaction of possibilities similarly as how classical physics contemplates the interaction of actualities. From this standpoint there is no need for invoking the {\it eigenstate-eigen\-value link} for both realms are independent. Thus, on the one hand, like in MI, only one way of the {\it if and only if} is required. The particular actualization (i.e., the measurement result) is a singular expression of the relation between the possible and actual realms, and is not considered as a physical interaction.  We could say that, like in MI,  the projection postulate is accepted but the collapse is denied. On the other hand, like in the MWI, every term in the superposition is interpreted as physically existent, however, there are no multiple (actual) worlds but rather a set of existent possibilities interacting in one single world.

\section{The MKS Theorem and Many Worlds}

The notion of possibility has been also investigated in relation to
the idea of possible worlds \cite{Lewis86}. Regarding QM, this
logical analysis has found an expression in the many worlds
interpretation \cite{DeWitt73}. In order to discuss this
notion of possibility within our own scheme, we have developed an
algebraic framework which allows us to analyze the modal aspects of
the MWI from a logical perspective \cite{DFR09b}.

According to the MWI all possibilities encoded in the wave function take place, but in
different worlds. When a measurement of a
physical magnitude ${\bf M}$ is performed and one of its possible
outcomes $a_1$ occurs, then in another world $a_2$ occurs, and in
some other world $a_3$ occurs, etc. In modal wording, suppose that
${\bf M}$ has associated a Boolean sublattice $W_{\bf M}$ of
$\mathcal{L}({\mathcal{H}})$. The projectors of the family $({\bf
P}_i)$ are identified as elements of $W_{\bf M}$. If a measurement
is performed and its result is $a_i$, this means that we can
establish a Boolean homomorphism $$v:W_{\bf M} \rightarrow {\bf 2}
\hspace{0.7cm}s.t. \hspace{0.2cm} v({\bf P}_i) = 1 $$ In a possible
world where $v({\bf P}_i) = 1$ we will have classical consequences.
Let us make precise the notion of {\it classical consequence} taking
into account modal extensions built from Boolean saturated
orthomodular lattices.

\begin{definition}
{\rm Let ${\cal L}^{\Diamond}$ be an arbitrary modal extension of
${\cal L}({ \mathcal{H}})$ and ${\bf P} \in {\cal L}({ \mathcal{H}})
$. Then $x \in \Diamond \mathcal{L}$ is said to be a {\it classical
consequence} of ${\bf P}$ iff for each Boolean sublattice $W$ in
${\cal L}^{\Diamond}$ (with ${\bf P}_i\in W$) and each Boolean
valuation $v:W \rightarrow {\bf 2}$, $v(x) = 1$ whenever $v({\bf
P}_i)=1$. }
\end{definition}

\noindent We denote by $Cons_{\mathcal{L}^\Diamond}({\bf P})$ the
set of classical consequences of $\mathcal{L}$. By Proposition
\cite[3.5]{DFR09b} we have that $Cons_{{\mathcal{L}}^\Diamond}({\bf
P}_i) = \{x\in \Diamond {\mathcal{L}}({\mathcal{H}}): \Diamond {\bf
P}_i \leq x \} $. The modal extension {\it does not depend} on the
valuation over the family $({\bf P}_i)$. Thus, it is clear that the
modal extension is independent of any possible world. Modal
extensions are simple algebraic extensions of an orthomodular
structure. Thus, when referring to a property ${\bf P}_i$, one can equivalently consider the classical consequences in the possible
world where $v({\bf P}_i) = 1$ or study the classical
consequences of $\Diamond {\bf P}_i$ before
the splitting. \\

Formally, MWI {\it maintains that in each respective
i-world, $v_i({\bf P}_i) = 1$ for each $i$}. Thus, a family of
valuations $(v_i({\bf P}_i) = 1)_i$ may be simultaneously
considered, each member being realized in each different $i$-world.
From an algebraic perspective, this would be equivalent to have a
family of pairs $\langle {\mathcal{L}}({\mathcal{H}}), v_i({\bf
P}_i) = 1 \rangle_i$, each pair being the orthomodular structure
${\mathcal{L}}({\mathcal{H}})$ with a distinguished Boolean
valuation $v_i$ over a spectral sub-algebra containing ${\bf P}_i$
such that $v_i({\bf P}_i) = 1$. In \cite{DFR09b}, we have shown that
the ${\cal OML}^\Diamond$ structure is able to capture  this fact in
terms of classical consequences. While MWI
considers a family of pairs $\langle {\mathcal{L}}({\mathcal{H}}),
v_i({\bf P}_i) = 1 \rangle_i$ for each possible i-world and the
classical consequences of $v_i({\bf P}_i) = 1$ in the $i$-world, the
${\cal OML}^\Diamond$ structure, by Proposition \cite[prop.
3.5]{DFR09b}, considers classical consequences of each $v_i({\bf
P}_i) = 1$ coexisting simultaneously in one and the same structure.
In fact, as a valuation $v: \Diamond \mathcal{L} \rightarrow {\bf
2}$ exists such that $v(\Diamond {\bf P}_i) = 1$ for each $i$, each
element $x\in \Diamond \mathcal{L}$ such that ${\bf P}_i \leq x$
necessarily satisfies $v(x)=1$.  In physical terms, this analysis shows that MWI talks
about possible propositions based on an orthomodular lattice without
taking into account the intrinsic features of the structure itself.
This has the consequence that, like in classical physics, in spite of the
wording about possibility that is present in the MWI, only actuality plays a role. Rather than using quantum possibilities, MWI restricts them to (classical) possibilities. One could say that all possibilities have become actual. Thus, the MKS theorem does not restrict the MWI scheme. Furthermore, there is no need of the projection postulate. MWI could be then considered as extending the MI proposed by Dieks to all terms of the multiple superpositions.\footnote{Private discussion, April 2013, Rio de Janeiro.}

\section{Conclusions}

In this paper we have discussed the characteristics of propositions referring to possibility within the framework of an
orthomodular lattice, in order to physically interpret the meaning and scope of our MKS theorem. In order to do so, we have distinguished the use of ``possibility'' in the classical and quantum formalisms. To escape from ambiguities in the relation between formalism and language, we have built a dictionary that clearly expresses the link between formal elements and physical concepts. The construction of the dictionary has also led us to the recognition of the independence between the realms of quantum possibility and actuality, in contradistinction to the classical case in which both possibility and actuality collapse. Furthermore, we have understood how the MKS theorem rules, through the constraints on actualization, the relation between both realms. Finally, we have analyzed the use of modality within the MWI, concluding that ---due to the fact that it does not directly confront the interpretation of quantum possibility--- it escapes both the KS and MKS theorems.

\section*{Acknowledgements}

We wish to thank an anonymous referee for his/her comments and recommendations on an earlier draft of this paper. C. de Ronde wishes to thank very specially Michiel Seevinck for proposing the need of a physical justification of the MKS theorem. This paper is greatly indebt for his questioning. He also wishes to thank Dennis Dieks for comments and suggestions on an earlier draft of this paper. This work was partially supported by the following grants: PIP 112-201101-00636, Ubacyt 2011/2014 635, FWO project G.0405.08 and FWO-research community W0.030.06. CONICET RES. 4541-12 (2013-2014).

\end{document}